\documentclass{article}
\usepackage{arxiv}

\usepackage{natbib}
\setcitestyle{round,authoryear}

\usepackage[utf8]{inputenc} 
\usepackage[T1]{fontenc}    
\usepackage{hyperref}       
\usepackage{url}            
\usepackage{booktabs}       
\usepackage{amsfonts}       
\usepackage{nicefrac}       
\usepackage{microtype}      
\usepackage{lipsum}
\usepackage{graphicx}

\usepackage{multicol}

\title{The Junk News Aggregator: Examining junk news posted on Facebook, starting with the 2018 US Midterm Elections}

\author{
  Dimitra (Mimie) Liotsiou\\
  Oxford Internet Institute\\
  University of Oxford\\
  Oxford, UK \\
  \texttt{mimie.liotsiou@oii.ox.ac.uk} \\
  \And 
  Bence Kollanyi \\
  Oxford Internet Institute\\
  University of Oxford\\
  Oxford, UK \\
  \texttt{bence.kollanyi@oii.ox.ac.uk} \\
   \And
 Philip N. Howard \\
  Oxford Internet Institute\\
  University of Oxford\\
  Oxford, UK \\
  \texttt{philip.howard@oii.ox.ac.uk} \\
}

\begin{document}
\maketitle

\begin{abstract}

In recent years, the phenomenon of online misinformation and junk news circulating on social media has come to constitute an important and widespread problem affecting public life online across the globe, particularly around important political events such as elections. At the same time, there have been calls for more transparency around misinformation on social media platforms, as many of the most popular social media platforms function as ``walled gardens," where it is impossible for researchers and the public to readily examine the scale and nature of misinformation activity as it unfolds on the platforms. In order to help address this, we present the Junk News Aggregator, a publicly available interactive web tool, which allows anyone to examine, in near real-time, all of the public content posted to Facebook by important junk news sources in the US. It allows the public to gain access to and examine the latest articles posted on Facebook (the most popular social media platform in the US and one where content is not readily accessible at scale from the open Web), as well as organise them by time, news publisher, and keywords of interest, and sort them based on all eight engagement metrics available on Facebook.  Therefore, the Aggregator allows the public to gain insights on the volume, content, key themes, and types and volumes of engagement received by content posted by junk news publishers, in near real-time, hence opening up and offering transparency in these activities as they unfold, at scale across the top most popular junk news publishers. In this way, the Aggregator can help increase transparency around the nature, volume, and engagement with junk news on social media, and serve as a media literacy tool for the public.


\end{abstract}


\section{Introduction}

The phenomenon of online misinformation and junk news circulating on social media has come to constitute an important problem affecting public life online across the globe, particularly around important political events such as elections. This phenomenon has become so widespread, and has continued to grow in scale and sophistication over the years, that it has been recognised as posing an important threat to the health of online political discourse and to democratic processes. At the same time, the volume, nature, and engagement with such content has often been difficult for researchers or the public to examine, due to the closed nature of many of the most popular social media platforms. 

In order to address this issue, in this paper we present the Junk News Aggregator(JNA)\footnote{\url{https://newsaggregator.oii.ox.ac.uk/}}, a website comprised of a suite of interactive web tools made publicly available, which allows the public to examine, in near real-time, all of the public content posted to Facebook by important junk news sources in the US.

Junk news sources are news sources that deliberately publish misleading, deceptive or incorrect information purporting to be real news about politics, economics or culture. This content includes ideologically extreme, hyper-partisan, or conspiratorial news and information, as well as various forms of propaganda.  

The goal in building the JNA was to help shed light on the problem of junk news on social media, to make this issue more transparent, and to help improve the public’s media literacy. We also aimed to help journalists, researchers, policy-makers, and social media platforms understand the impact of junk news on public life. 

We first launched the JNA on November 1st 2018, in the lead-up to the 2018 US midterm elections, as the previous US elections in 2016 were recognised for the online misinformation that was circulated, and misinformation in relation to US politics has far from ceased since. This tool allows the public to gain access to and examine the latest articles posted on Facebook, the most popular social media platform in the US and one where content is not readily accessible at scale from the open Web, as well as filter them by time (from the last hour to one month in the past), news publisher, and keywords of interest, and sort them based on all eight engagement metrics available on Facebook. 

The JNA offers three tools, each of which offers a different type of window into this activity: a full interactive Explorer with all the filtering and sorting functionality, and a Daily Visual Grid\footnote{\url{https://newsaggregator.oii.ox.ac.uk/index.php\#visual_jna}} of all images in this activity which links to the full aggregator, as well as a static Daily Top-10 List\footnote{\url{https://newsaggregator.oii.ox.ac.uk/top10.php}} showing the posts that received the most engagement in the last day. Therefore, the Aggregator allows the public to gain insights on the volume, content, key themes, and types and amount of engagement received by content posted by junk news publishers, in near real-time, hence opening up and offering transparency in these activities, at scale across the top most popular junk news publishers. In this way, the Aggregator can help increase transparency around the nature, volume, and engagement with junk news on social media, and serve as a media literacy tool for the public. 

The JNA takes as an input a list of websites and their associated public Facebook Pages, so it can be deployed to track any set of public Facebook Pages of interest. In the future, we plan to extend the JNA so that it can track junk news related to future elections or other important political events, potentially tracking multiple such events at the same time. 


The JNA is an interactive web tool for exploring junk news stories posted on Facebook in the lead-up to the US midterms, and it enables website visitors to examine and parse content posted by junk news sources on Facebook in near real-time. It make visible the depth of the junk news problem, displaying the quantity and the content of junk news, as well as the levels of engagement with it. Junk news content can be sorted by time and by engagement numbers, as well as via keyword search (such as for a candidate, district, or specific issue). It also offers a visual overview (the Daily Visual Grid), and a top-10 snapshot of the day’s most engaged-with junk news (the Daily Top-10 List).

This paper describes the JNA -- how it was built, what data it displays, and the methodology based on which the data was collected. The paper is strucutred as follows: the first section presents relevant background, discussing what junk news is, and the context of online misinformation. The second section proceeds to describe the methodology used to collect data for the JNA. Next, the paper discusses how the collected data is presented on the JNA website, discussing each of the three tools made available for the user to examine, organise, and explore junk news posts on Facebook. The paper ends with a brief concluding remarks on the goals and contributions of the JNA.



\section{Background}


\subsection{What is ``junk news"?}
Per the definitions in \cite{bolsover2018chinese}, \cite{gallacher2018junk}, \cite{howard2017junk}, \cite{howard2018algorithms}, and \cite{woolley2017exec}, the term ``junk news" refers to various forms of propaganda and ideologically extreme, hyper-partisan, or conspiratorial political news and information. The term includes news publications that present verifiably false content as factual news. This content includes propagandistic, ideologically extreme, hyper-partisan, or conspiracy-oriented news and information. Frequently, attention-grabbing techniques are used, such as lots of pictures, moving images, excessive capitalization, personal attacks, emotionally charged words and pictures, populist generalizations, and logical fallacies. It presents commentary as news. The term refers to a news source overall, i.e. based on content that is typically published by a source, rather than referring to an individual article. Further context on junk news and online misinformation can be found in the next section.

The five criteria listed in Table~\ref{tbl:jn_criteria} are used to determine whether a website is a source of junk news. If a website satisfies the majority, i.e. three or more, of these five criteria, it is considered a source of junk news.

\begin{table}[!ht]
\caption{Junk News Criteria}
\label{tbl:jn_criteria}
\centering
  \begin{tabular}{p{5cm} p{9cm}}
  \textbf{Criterion (Abbreviation)}                            &  \textbf{Full Description}    \\ \hline
  Professionalism (P)                                 & These sources do not employ standards and best practices of professional journalism. They refrain from providing clear information about real authors, editors, publishers and owners. They lack transparency and accountability, and do not publish corrections on debunked information.                           \\
  Style (S)                                           & These sources use emotionally driven language with emotive expressions, hyperbole, ad hominem attacks, misleading headlines, excessive capitalization, unsafe generalizations and logical fallacies, moving images, and lots of pictures and mobilizing memes.                                                      \\
  Credibility (Cr)                                    & These sources rely on false information and conspiracy theories, which they often employ strategically. They report without consulting multiple sources and do not fact-check. Sources are often untrustworthy and standards of production lack reliability.                                                        \\
  Bias (B), Left-wing bias (LB), Right-wing bias (RB) & Reporting in these sources is highly biased, ideologically skewed or hyper-partisan, and news reporting frequently includes strongly opinionated commentary and inflammatory viewpoints.                                                                                                                            \\
  Counterfeit (Ct)                                    & These sources mimic established news reporting. They counterfeit fonts, branding and stylistic content strategies. Commentary and junk content is stylistically disguised as news, with references to news agencies and credible sources, and headlines written in a news tone with date, time and location stamps. \\
  Aggregator of junk news (JN AGGR)                   & These sources aggregate other sources that are themselves deemed to be junk news sources based on the above criteria.                                                                                                                                                                                              
  \end{tabular}
\end{table}

\subsection{Online misinformation and content typologies}
The above criteria for junk news are part of a typology for classifying news content (a recent version of this typology can be found in \cite{marchal2018polarization}). This section motivates and contextualises the need for such a typology. 

Following the highly contentious 2016 US presidential election, there has been a growing body of empirical work demonstrating how large volumes of misinformation can circulate over social media during critical moments of public life \citep{allcott2017social, del2016spreading, vosoughi2018spread}. Scholars have argued that the spread of ``computational propaganda" sustained by social media algorithms can negatively impact democratic discourse and disrupt digital public spheres \citep{bradshaw2018why, howard2016political, persily20172016, tucker2017liberation, wardle2017information}. Indeed, both social network infrastructure and user behaviors provide capacities and constraints for the spread of computational propaganda \citep{bradshaw2018why, flaxman2016filter, marwick2017media, pariser2011filter, wu2017attention}. Yet, the body of work that is devoted to conceptualizing misinformation phenomena faces a number of epistemological and methodological challenges, has remained fragmentary, is ambiguous at best and lacks a common vocabulary \cite{boyd2017google,fletcher2018measuring, wardle2017information}. Terminology on misinformation has become highly contentious, constantly being weaponised by politically motivated actors to discredit media reporting \citep{neudert2017computational,woolley2018computational}.

There have been a few attempts to systematically operationalize fake news as a concept. The primary challenge is that it is impossible to evaluate the amount of fact checking that goes into a particular piece of writing at a scale sufficient for saying something general about the trends on a social media platform. Most researchers—and indeed most citizens—have manual heuristics for evaluating sources that deliberately publish or aggregate misleading, deceptive or incorrect information purporting to be real news about politics, economics or culture. For different social media users there are different ways of evaluating the qualities of news and political information on social media.

Typologies in political communication research have been useful for frame analysis for the study of news, or the organization of event-based datasets in which media accounts provide the primary features for important incidents  \citep{althaus2001using, erickson2007case}. Even before social media, scholars used such methods to expose the ways in which sensational news organizations used human interest frames, while serious news organizations used responsibility and conflict frames \citep{semetko2000framing}. In order to understand what users were actually sharing over social media, we have developed a typology of political news and information. 

Content typologies are of central importance to the study of political communication, and for many years the broad categories and subcategories of political news and information have remained widely accepted by researchers, though of course there is debate over how transportable such traditional categories are to new media political communication \cite{earl2004use,karlsson2016content}. However, the recent attention and debate over the effects of junk news in the media ecosystem have forced researchers—especially those working in political communication and social media—to re-evaluate the production models, normative values, and ideational impact of political news and information of social media with new categories and definitions that are actually grounded in the content being shared over social media \cite{tandoc2018defining}. Currently, the debate lacks a grounded and comparative framework of the types of information that circulate on social media, and remains detached from evidence of social media sharing behavior. The rigorously composed typologu we use advances this debate, and is based on a focused, cross-case comparison of key political events in the US.

The typology we use proceeds from a recognition that what users consume and share over social media in their political conversations is not simply news, but could include a wide variety of sources for political news and information, including user-generated content, conspiratorial alternative media outlets, and entertainment outlets. Indeed, there is significant research that humorous content is a staple of information sharing in contemporary political communication \cite{becker2012comedy, becker2010sizing, moy2005priming}. Given the lack of guidance in the existing literature on what information users are sharing, a grounded and iterative method of cataloguing content and evaluation is especially relevant.

\section{Methodology}


The JNA offers the public a suite of three tools, which enable one to track in near real-time the junk news content being posted and engaged with on Facebook, and to filter this information by time, keyowrd, and by each of the eight engagement types available on Facebook.

The JNA queries Facebook every hour on the hour, using the public Facebook Graph API, and collects only public data. No sensitive, personal or user-identifying data is collected. Specifically, it queries a list of Public Facebook Pages of specific junk news publishers. These junk news publishers were chosen because their web content was found to be particularly frequently discussed in social media conversations relating to the 2018 US midterm elections. The list of junk news publishers was assembled and utilised in the JNA according to the following sequence of steps:

\begin{enumerate}

\item Select Twitter hashtags relevant to the 2018 US midterm elections. Ensure that these hashtags relate to this specific election and not to other 2018 elections around the world, and ensure the list of hashtags is balanced, in the sense of covering both right and left-leaning hashtags.

\item Use the Twitter Streaming API, to get any other hashtags that are mentioned together with the hashtags in our list: get all tweets mentioning any of the hashtags in our list, and from these tweets, get other hashtags mentioned in those tweets. This snowball sampling of hashtags results in an expanded list of hashtags. This snowball sampling for the hashtags happened from September 15 to September 19, 2018. The list of hashtags used is shown in Table \ref{tbl:hashtags}. 


\item Using this expanded hashtag list, use the Twitter Streaming API to retrieve all English-language tweets (including retweets and quote tweets) mentioning any of these hashtags. This data collection resulted in 2,541,544 tweets posted in the period September 21 to September 30, 2018.

\item Out of this set of tweets, get all the URLs they mention. Keep only the base URL (i.e. not a specific article's URL but the homepage URL), and count how many times each base URL was mentioned.

\item Give this list of URLs to exparts trained in the US context (trained human annotators or ``coders"), to classify all URLs into categories of news and political content, using a grounded typology \cite{woolley2018computational}. For the junk news category, there exist five criteria, so if a news source satisfies at least three of these five criteria, it gets classified under this category. To train our team of US experts to categorize sources of political news and information according to our grounded typology, we established a rigorous training system. For the analysis of the 2018 US midterms, we worked with a team of three coders. Each source was triple-coded. Any conflicting decision was thoroughly discussed between coders to achieve consensus. In the event that consensus was not achieved, an executive team of three other highly experienced coders reviewed the source and made a final coding decision.

\item For every website in the Junk News list that has been shared on Twitter, identify their Facebook page, if they have any. In order to establish that a given Facebook page corresponds to a given website, it is required that either the website explicitly lists this Facebook page as theirs, and/or that the given Facebook page lists under its 'Website' field this particular website.

\item Out of all junk news sites, keep only the ones that have a Facebook page. Out of those, keep only the top 50 most shared ones on Twitter (due to the Facebook Graph API's rate limits, not all of them can be tracked). These 50 junk news sites, along with the typology criteria which they satisfy (due to which they get classified as junk news), can be found in Table \ref{tbl:jn_sources_criteria}, where the explanation of the code (abbreviation) used for each criterion is described above.


\begin{table}[!ht]
\caption{Twitter hashtags used}
\label{tbl:hashtags}
\centering
\begin{tabular}{lll}      
2018midterms       & getoutthevote     & voteblue                  \\
bluetsunami        & gop               & voteblue2018              \\
bluetsunami2018    & midtermelections  & votebluetosaveamerica     \\
bluewave           & midterms          & votedemout                \\
bluewave2018       & midterms2018      & votedemsout               \\
bluewavecoming2018 & rednationrising   & votegopout                \\
bluewaveiscoming   & redwave           & votered                   \\
dem                & redwave2018       & votered2018               \\
democrat           & redwaverising     & voteredtosaveamerica      \\
democrats          & redwaverising2018 & walkaway                  \\
dems               & republican        & walkawayfromdemocrats2018 \\
earlyvoting        & republicans       & whenweallvote             \\
flipitblue         & uniteblue         &                          
\end{tabular}
\end{table}


\begin{table}[!ht]
\caption{Junk News Sources and Relevant Criteria Violated}
\label{tbl:jn_sources_criteria}
\centering
\begin{tabular}{lllll}
\textbf{Source}  &  \multicolumn{4}{l}{\textbf{Junk News Criteria Satisfied}} \\ \hline     
breitbart.com             & RB            & S  & Cr      & P       \\
thegatewaypundit.com      & RB            & S  & Cr      & P       \\
libertyheadlines.com      & RB            & S  & Cr     &          \\
theblacksphere.net        & RB            & Cr & S       & P       \\
dailywire.com             & RB            & S  & Cr     &          \\
thefederalist.com         & RB            & S  & Cr     &          \\
rawstory.com              & LB            & Cr & P      &          \\
thedailydigest.org        & CR            & Ct & RB      & P       \\
lifenews.com              & RB            & S  & Cr      & P       \\
infowars.com              & RB            & S  & Cr      & P       \\
dailycaller.com           & RB            & S  & P      &          \\
zerohedge.com             & Cr            & P  & S      &          \\
barenakedislam.com        & RB            & S  & Cr      & P       \\
pjmedia.com               & P             & Ct & Cr     &          \\
americanthinker.com       & RB            & S  & Cr      & Ct      \\
newrightnetwork.com       & Ct            & S  & RB     &          \\
gellerreport.com          & RB            & S  & Cr      & Ct      \\
davidharrisjr.com         & RB            & S  & P      &          \\
theoldschoolpatriot.com   & S             & RB & Cr     &          \\
100percentfedup.com       & RB            & S  & Cr      & P       \\
committedconservative.com & RB            & S  & Cr      & P       \\
truthfeednews.com         & P             & Ct & Cr      & RB      \\
michaelsavage.com         & RB            & S  & Cr     &          \\
bigleaguepolitics.com     & RB            & S  & Cr     &          \\
cnsnews.com               & RB            & S  & Cr      & P       \\
truepundit.com            & RB            & S  & P      &          \\
thepoliticalinsider.com   & P             & Cr & Ct      & RB      \\
hotair.com                & RB            & S  & Cr     &          \\
lifezette.com             & RB            & S  & Cr      & P       \\
canadafreepress.com       & RB            & S  & Cr      & P       \\
shareblue.com             & LB            & P  & CT     &          \\
wnd.com                   & Ct            & RB & P      &          \\
bizpacreview.com          & RB            & S  & P      &          \\
rushlimbaugh.com          & RB            & P  & S      &          \\
theblaze.com              & P             & RB & Ct     &          \\
frontpagemag.com          & RB            & S  & Cr      & P       \\
redstate.com              & RB            & S  & Ct      & P       \\
palmerreport.com          & LB            & P  & S      &          \\
chicksonright.com         & RB            & S  & Cr     &          \\
nworeport.me              & S             & RB & P       & Ct      \\
en-volve.com              & RB            & S  & Cr     &          \\
magaoneradio.net          & RB            & S  & P       & JN AGG  \\
twitchy.com               & S             & RB & P      &          \\
naturalnews.com           & RB            & S  & Cr      & P       \\
westernfreepress.com      & P             & S  & RB     &          \\
legalinsurrection.com     & RB            & S  & P       & Ct      \\
conservativedailypost.com & RB            & S  & Cr      & P       \\
therightscoop.com         & RB            & S  & P      &          \\
conspiracydailyupdate.com & Cr            & S  & JN AGG &         
\end{tabular}
\end{table}

\item Out of these 50 junk news sites, for the public Facebook page of each, retrieve the public posts authored by them and some of the post metadata (including aggregate-level engagement numbers): every hour on the hour, get all posts authored by these pages, not including any names of people who engage with these posts, but rather only numbers of engagements (reactions), for all eight types of engagement that Facebook makes available: Share, Comment, Like, Love, Haha, Wow, Sad, Angry. Write these posts to a database.

\item For the Junk News Aggregator site, retrieve data from this database of public Facebook posts by junk news publishers, and allow site visitors to explore, sort, and filter this data by time, engagement numbers, and keywords.
\end{enumerate}

We note that, for each Facebook post collected, the displayed engagement numbers were last updated at most an hour after the post was posted, as, due to the Facebook API's data limits, we cannot update those again later. But the link to the Facebook post is provided, so you can click on that to see current engagement levels for this post.


\section{The JNA website}

This section describes the three tools available on the JNA website: the interactive Explorer, as well as two simpler and less interactive tools that show the top posts of the last 24 hours: the Daily Top-10 List, and the Daily Visual Grid.

\subsection{The Daily Visual Grid}
\begin{figure}[!ht]
  \centering
  \includegraphics[width=120mm]{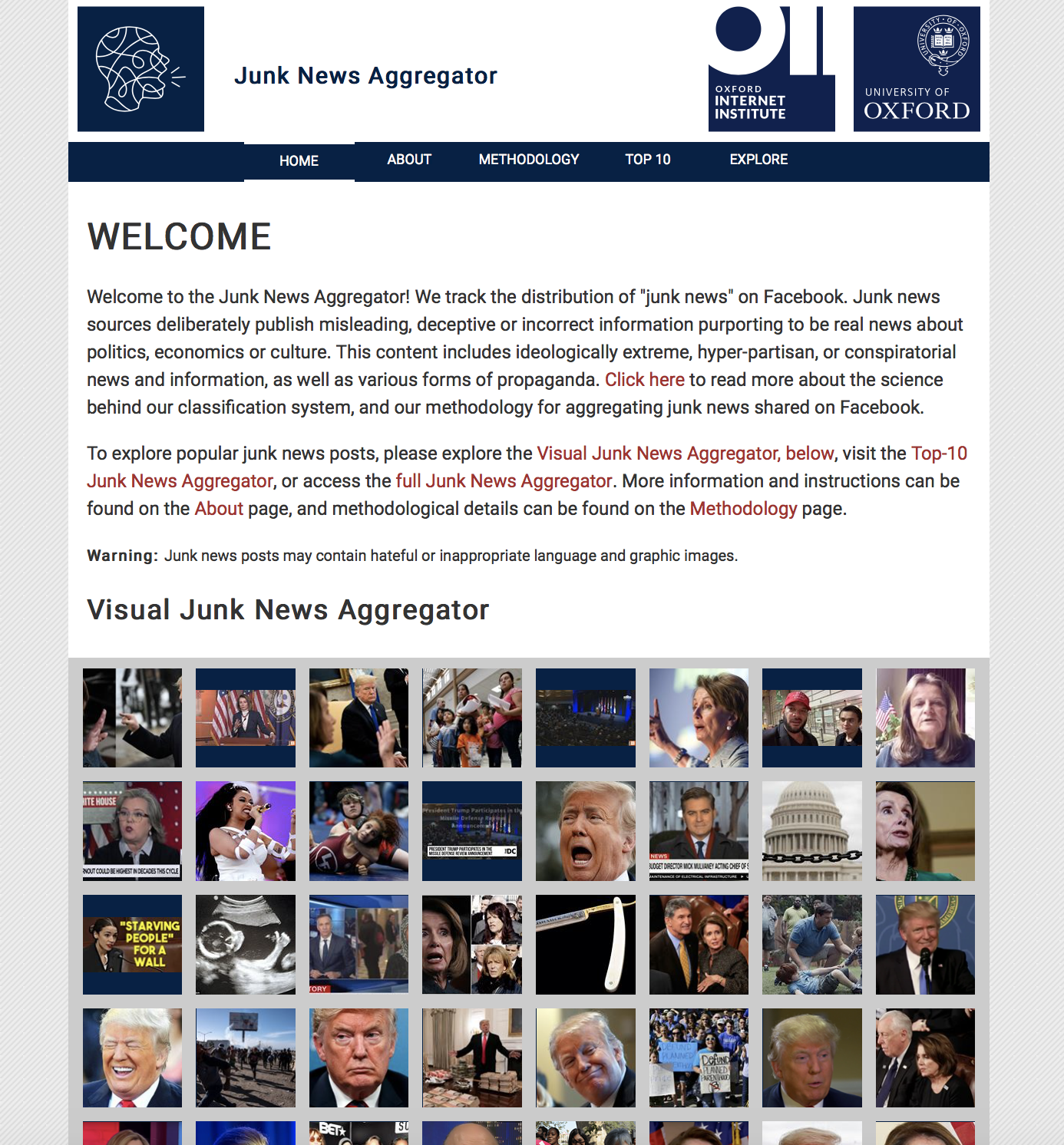} 
  \caption{Screenshot of the Daily Visual Grid} 
  \label{fig:visual}
\end{figure}

On the JNA homepage, one finds an interactive 16x16 image grid\footnote{\url{https://newsaggregator.oii.ox.ac.uk/index.php\#visual_jna}}. This is the Daily Visual Grid, showing square images from the top-256 junk news posts with the most age-adjusted total engagements in the last 24 hours. Figure \ref{fig:visual} shows a screenshot of this Daily Visual Grid.  

This Daily Visual Grid updates every 24 hours at 5pm ET, and shows only the top-256 most engaged with Facebook posts that were posted in this 24-hour period by junk news sources in a 16x16 grid. These are the top posts based on age-adjusted total engagements (the ``age-adjusted total: All" metric). Each image in the grid corresponds to a junk news Facebook post. Hovering over an image reveals a pop-up showing more information about the relevant post: the Facebook Page that posted it, the time and date posted, the text in the post, and engagement numbers. Clicking on an image takes the user to the Explorer, where one can explore junk news in greater detail.


\subsection{The Daily Top-10 List}
From the homepage, one can navigate to the Daily Top-10 List\footnote{\url{https://newsaggregator.oii.ox.ac.uk/top10.php}}. This is a smaller and simpler tool than the interactive Explorer. It uses the same data as the full Explorer, but it queries this data less frequently, once every 24 hours, at 5pm ET, and shows only the top 10 most engaged-with Facebook posts that were posted in this 24-hour period by junk news sources, in terms of the overall age-adjusted total engagements these posts received. A post's overall age-adjusted total engagements is the sum of all engagements received by this post (the number of Likes + Comments + Shares + Love reactions + Haha reactions + Wow reactions + Angry reactions + Sad reactions) divided by the post's age in seconds, where a post's age equals the time when the post was retrieved from Facebook minus the time when the post was posted to Facebook, with this age measured here in seconds.

\begin{figure}[!ht]
  \centering
  \includegraphics[width=120mm]{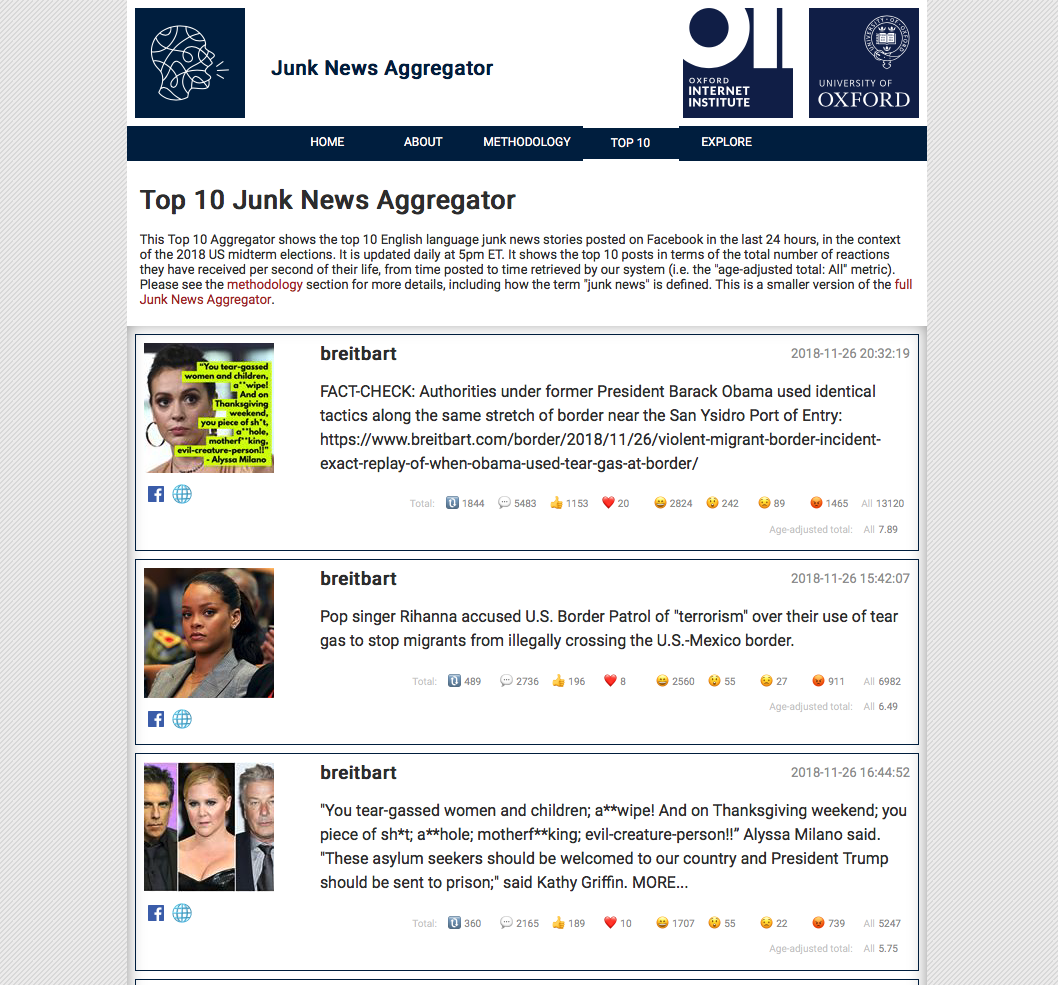} 
  \caption{Screenshot of the Daily Top-10 List} 
  \label{fig:top10}
\end{figure}

A screenshot of the Daily Top-10 List is shown in Figure \ref{fig:top10}. 

\subsection{The Explorer}
\begin{figure}[!ht]
	\centering
	\includegraphics[width=120mm]{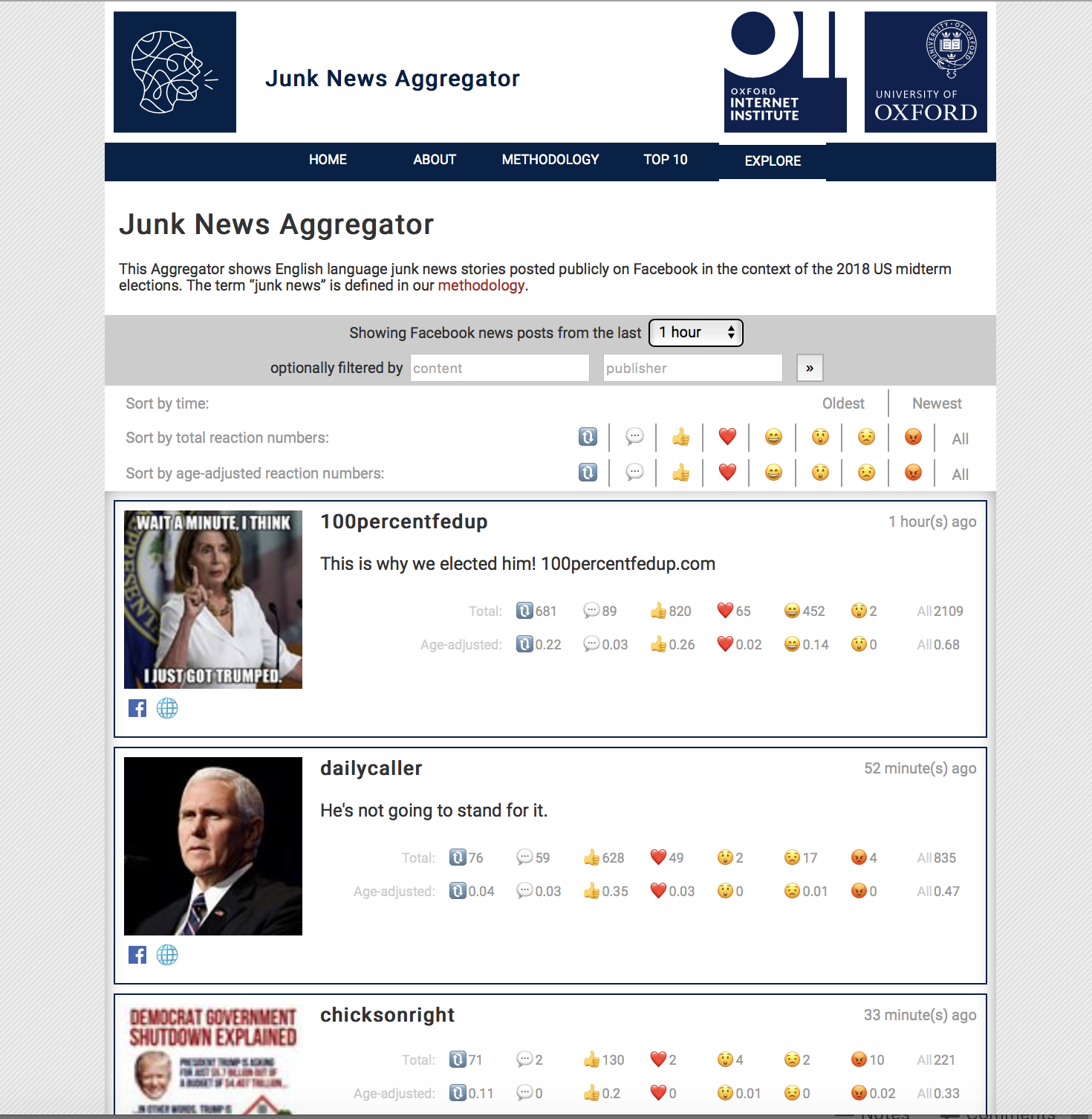} 
	\caption{Screenshot of the Explorer} 
	\label{fig:full_jna}
\end{figure}

From the JNA homepage, one can navigate to the full interactive Explorer\footnote{\url{https://newsaggregator.oii.ox.ac.uk/news/app/}}, either by clicking on any image in the Visual Grid, or by clicking on the ``Explore" tab. It shows Facebook posts posted by junk news outlets on their public Facebook page. It shows up-to-the-hour posts, going as far back as a month into the past. A screenshot of the Explorer is shown in Figure \ref{fig:full_jna}. 


The user can filter posts based on how long ago they were posted (e.g. 1 hour ago, 2 hours ago, etc.), and also based on keywords and the publisher name. The user can also sort posts, not only by when they were posted (newest/ oldest), but also by how many engagements (or reactions) they received, for each of the eight post engagement types available on Facebook (Likes, Comments, Shares, and the five emoji reactions: Love, Haha, Wow, Angry, Sad), and by the sum of all engagements across all eight metrics ("All"). 

In addition, the user can sort posts by the age-adjusted version of each of the engagement types (number of engagements divided by the post's age in seconds), which shows the number of engagements a post received per second of its life on Facebook (up to the point it was retrieved by the JNA system). Since the JNA system only queries Facebook once an hour (due to Facebook API's rate limits, this cannot be done more frequently), this accounts for the age of the post at the time Facebook was queried, and offers a more appropriate measure for comparing and sorting posts based on engagement numbers.

\section{Conclusion}
Given the proliferation of misinformation and junk news on social media platforms around important election events, we have built the Junk News Aggregator (JNA), to enable the public to transparently and systematically examine and parse junk news on Facebook in near real-time, hence contributing towards transparency around junk news on social media. This is achieved through this interactive web app, by making visible the quantity and the content that US junk news published publicly on Facebook, as well as the levels of engagement with it. This effort started shortly before the 2018 US midterm elections, and continues to this day. The JNA website is comprised of three tools, each offering a different window into the online activity of these junk news sources: a Daily Visual Grid, focusing on the visual media content that these sources employ, which leads to the interactive Explorer, which displays the full content of each Facebook post and offers more extensive filtering and sorting functionality and goes up to a month back (temporal, keyword-based, and engagement-based, for all eight engagement actions available for Facebook posts), and the Daily Top-10 List, offering a snapshot of the day's top engaged-with Facebook posts from these junk news publishers. In this manner, the JNA is the first publicly-available tool that offers insights into the content publicly uploaded to Facebook by junk news sources, contributing towards increasing transparency around the nature, volume, and engagement with junk news on social media, and aiming to serve as a media literacy tool for the public.


\section{Acknowledgements}
This work is funded by the European Research Council through the grant “Computational Propaganda: Investigating the Impact of Algorithms and Bots on Political Discourse in Europe,” Proposal 648311, 2015-2020, Philip N. Howard, Principal Investigator, and the grant "Restoring Trust in Social Media Civic Engagement," Proposal Number: 767454, 2017-2018, Philip N. Howard, Principal Investigator. We are grateful for additional support from the Open Society Foundation and Ford Foundation. Project activities were approved by the University of Oxford’s Research Ethics Committee, CUREC OII C1A 15 044, C1A 17 054. Any opinions, findings, and conclusions or recommendations expressed in this material are those of the researchers and do not necessarily reflect the views of the funders, or the University of Oxford.

We thank Shaun A. Noordin, Adham Tamer and John Gilbert of the OII, and Mike Antcliffe of Achromatic Security, for their valuable web development and security testing work on this project. Thanks also to Lisa-Maria Neudert and Vidya Narayanan, and to the rest of the COMPROP team, for their very helpful suggestions, feedback and support during the development of this project.

\bibliographystyle{plainnat}
\bibliography{jna}

\begin{thebibliography}{31}
\providecommand{\natexlab}[1]{#1}
\providecommand{\url}[1]{\texttt{#1}}
\expandafter\ifx\csname urlstyle\endcsname\relax
  \providecommand{\doi}[1]{doi: #1}\else
  \providecommand{\doi}{doi: \begingroup \urlstyle{rm}\Url}\fi

\bibitem[Allcott and Gentzkow(2017)]{allcott2017social}
Hunt Allcott and Matthew Gentzkow.
\newblock Social media and fake news in the 2016 election.
\newblock \emph{Journal of Economic Perspectives}, 31\penalty0 (2):\penalty0
  211--36, 2017.

\bibitem[Althaus et~al.(2001)Althaus, Edy, and Phalen]{althaus2001using}
Scott~L Althaus, Jill~A Edy, and Patricia~F Phalen.
\newblock Using substitutes for full-text news stories in content analysis:
  Which text is best?
\newblock \emph{American Journal of Political Science}, pages 707--723, 2001.

\bibitem[Becker(2012)]{becker2012comedy}
Amy~B Becker.
\newblock Comedy types and political campaigns: The differential influence of
  other-directed hostile humor and self-ridicule on candidate evaluations.
\newblock \emph{Mass Communication and Society}, 15\penalty0 (6):\penalty0
  791--812, 2012.

\bibitem[Becker et~al.(2010)Becker, Xenos, and Waisanen]{becker2010sizing}
Amy~B Becker, Michael~A Xenos, and Don~J Waisanen.
\newblock Sizing up the daily show: Audience perceptions of political comedy
  programming.
\newblock \emph{Atlantic Journal of Communication}, 18\penalty0 (3):\penalty0
  144--157, 2010.

\bibitem[Bolsover and Howard(2018)]{bolsover2018chinese}
Gillian Bolsover and Philip~N Howard.
\newblock Chinese computational propaganda: automation, algorithms and the
  manipulation of information about {C}hinese politics on {T}witter and
  {W}eibo.
\newblock \emph{Information, Communication \& Society}, pages 1--18, 2018.

\bibitem[boyd(2017)]{boyd2017google}
d~boyd.
\newblock Google and {F}acebook can’t just make fake news disappear.
\newblock \emph{WIRED}, 2017.
\newblock URL
  \url{https://www.wired.com/2017/03/google-and-facebook-cant-just-make-fake-news-disappear/}.

\bibitem[Bradshaw and Howard(2018)]{bradshaw2018why}
Samantha Bradshaw and Philip~N Howard.
\newblock Why does junk news spread so quickly across social media? algorithms,
  advertising and exposure in public life.
\newblock \emph{Knight Foundation, Working Paper}, January 2018.
\newblock URL
  \url{https://kf-site-production.s3.amazonaws.com/media_elements/files/000/000/142/original/Topos_KF_White-Paper_Howard_V1_ado.pdf}.

\bibitem[Del~Vicario et~al.(2016)Del~Vicario, Bessi, Zollo, Petroni, Scala,
  Caldarelli, Stanley, and Quattrociocchi]{del2016spreading}
Michela Del~Vicario, Alessandro Bessi, Fabiana Zollo, Fabio Petroni, Antonio
  Scala, Guido Caldarelli, H~Eugene Stanley, and Walter Quattrociocchi.
\newblock The spreading of misinformation online.
\newblock \emph{Proceedings of the National Academy of Sciences}, 113\penalty0
  (3):\penalty0 554--559, 2016.

\bibitem[Earl et~al.(2004)Earl, Martin, McCarthy, and Soule]{earl2004use}
Jennifer Earl, Andrew Martin, John~D McCarthy, and Sarah~A Soule.
\newblock The use of newspaper data in the study of collective action.
\newblock \emph{Annu. Rev. Sociol.}, 30:\penalty0 65--80, 2004.

\bibitem[Erickson and Howard(2007)]{erickson2007case}
Kris Erickson and Philip~N Howard.
\newblock A case of mistaken identity? news accounts of hacker, consumer, and
  organizational responsibility for compromised digital records.
\newblock \emph{Journal of Computer-Mediated Communication}, 12\penalty0
  (4):\penalty0 1229--1247, 2007.

\bibitem[Flaxman et~al.(2016)Flaxman, Goel, and Rao]{flaxman2016filter}
Seth Flaxman, Sharad Goel, and Justin~M Rao.
\newblock Filter bubbles, echo chambers, and online news consumption.
\newblock \emph{Public Opinion Quarterly}, 80\penalty0 (S1):\penalty0 298--320,
  2016.

\bibitem[Fletcher et~al.(2018)Fletcher, Cornia, Graves, and
  Nielsen]{fletcher2018measuring}
Richard Fletcher, Alessio Cornia, Lucas Graves, and Rasmus~Kleis Nielsen.
\newblock Measuring the reach of “fake news” and online disinformation in
  {E}urope.
\newblock \emph{Reuters Institute Factsheet}, 2018.

\bibitem[Gallacher et~al.(2018)Gallacher, Barash, Howard, and
  Kelly]{gallacher2018junk}
John~D Gallacher, Vlad Barash, Philip~N Howard, and John Kelly.
\newblock Junk news on military affairs and national security: Social media
  disinformation campaigns against {US} military personnel and veterans.
\newblock \emph{ArXiv preprint arXiv:1802.03572}, 2018.

\bibitem[Howard and Woolley(2016)]{howard2016political}
Philip~N Howard and Samuel Woolley.
\newblock Political communication, computational propaganda, and autonomous
  agents: Introduction.
\newblock \emph{International Journal of Communication}, 10:\penalty0
  4882--4890, 2016.

\bibitem[Howard et~al.(2017)Howard, Bolsover, Kollanyi, Bradshaw, and
  Neudert]{howard2017junk}
Philip~N Howard, Gillian Bolsover, Bence Kollanyi, Samantha Bradshaw, and
  Lisa-Maria Neudert.
\newblock Junk news and bots during the {US} election: What were {M}ichigan
  voters sharing over {T}witter.
\newblock \emph{Computational Propaganda Research Project, Oxford Internet
  Institute, Data Memo, 2017.1}, 2017.

\bibitem[Howard et~al.(2018)Howard, Woolley, and Calo]{howard2018algorithms}
Philip~N Howard, Samuel Woolley, and Ryan Calo.
\newblock Algorithms, bots, and political communication in the {US} 2016
  election: The challenge of automated political communication for election law
  and administration.
\newblock \emph{Journal of Information Technology \& Politics}, 15\penalty0
  (2):\penalty0 81--93, 2018.

\bibitem[Karlsson and Sj{\o}vaag(2016)]{karlsson2016content}
Michael Karlsson and Helle Sj{\o}vaag.
\newblock Content analysis and online news: epistemologies of analysing the
  ephemeral web.
\newblock \emph{Digital Journalism}, 4\penalty0 (1):\penalty0 177--192, 2016.

\bibitem[Marchal et~al.(2018)Marchal, Neudert, Kollanyi, and
  Howard]{marchal2018polarization}
Nahema Marchal, Lisa-Maria Neudert, Bence Kollanyi, and Philip~N Howard.
\newblock Polarization, partisanship and junk news consumption on social media
  during the 2018 us midterm elections.
\newblock \emph{Computational Propaganda Research Project, Oxford Internet
  Institute, Data Memo, 2018.5}, 2018.

\bibitem[Marwick and Lewis(2017)]{marwick2017media}
Alice Marwick and Rebecca Lewis.
\newblock Media manipulation and disinformation online.
\newblock \emph{New York: Data \& Society Research Institute}, 2017.

\bibitem[Moy et~al.(2005)Moy, Xenos, and Hess]{moy2005priming}
Patricia Moy, Michael~A Xenos, and Verena~K Hess.
\newblock Priming effects of late-night comedy.
\newblock \emph{International Journal of Public Opinion Research}, 18\penalty0
  (2):\penalty0 198--210, 2005.

\bibitem[Neudert(2017)]{neudert2017computational}
Lisa-Maria Neudert.
\newblock Computational propaganda in {G}ermany: A cautionary tale.
\newblock Technical report, 2017.

\bibitem[Pariser(2011)]{pariser2011filter}
Eli Pariser.
\newblock \emph{The filter bubble: How the new personalized web is changing
  what we read and how we think}.
\newblock Penguin, 2011.

\bibitem[Persily(2017)]{persily20172016}
Nathaniel Persily.
\newblock The 2016 {US} election: Can democracy survive the internet?
\newblock \emph{Journal of Democracy}, 28\penalty0 (2):\penalty0 63--76, 2017.

\bibitem[Semetko and Valkenburg(2000)]{semetko2000framing}
Holli~A Semetko and Patti~M Valkenburg.
\newblock Framing european politics: A content analysis of press and television
  news.
\newblock \emph{Journal of communication}, 50\penalty0 (2):\penalty0 93--109,
  2000.

\bibitem[Tandoc~Jr et~al.(2018)Tandoc~Jr, Lim, and Ling]{tandoc2018defining}
Edson~C Tandoc~Jr, Zheng~Wei Lim, and Richard Ling.
\newblock Defining “fake news” a typology of scholarly definitions.
\newblock \emph{Digital Journalism}, 6\penalty0 (2):\penalty0 137--153, 2018.

\bibitem[Tucker et~al.(2017)Tucker, Theocharis, Roberts, and
  Barber{\'a}]{tucker2017liberation}
Joshua~A Tucker, Yannis Theocharis, Margaret~E Roberts, and Pablo Barber{\'a}.
\newblock From liberation to turmoil: social media and democracy.
\newblock \emph{Journal of Democracy}, 28\penalty0 (4):\penalty0 46--59, 2017.

\bibitem[Vosoughi et~al.(2018)Vosoughi, Roy, and Aral]{vosoughi2018spread}
Soroush Vosoughi, Deb Roy, and Sinan Aral.
\newblock The spread of true and false news online.
\newblock \emph{Science}, 359\penalty0 (6380):\penalty0 1146--1151, 2018.

\bibitem[Wardle and Derakhshan(2017)]{wardle2017information}
Claire Wardle and Hossein Derakhshan.
\newblock Information disorder: Toward an interdisciplinary framework for
  research and policymaking.
\newblock \emph{Council of Europe report, DGI (2017)}, 9, 2017.

\bibitem[Woolley and Howard(2017)]{woolley2017exec}
Samuel Woolley and Philip~N Howard.
\newblock Computational propaganda: Executive summary.
\newblock \emph{Computational Propaganda Research Project, Oxford Internet
  Institute, Working Paper 2017.11}, 2017.

\bibitem[Woolley and Howard(2018)]{woolley2018computational}
Samuel Woolley and Philip~N Howard.
\newblock \emph{Computational Propaganda: Political Parties, Politicians, and
  Political Manipulation on Social Media}.
\newblock Oxford University Press, 2018.

\bibitem[Wu(2017)]{wu2017attention}
Tim Wu.
\newblock \emph{The attention merchants: The epic scramble to get inside our
  heads}.
\newblock Vintage, 2017.

\end{thebibliography}

\end{document}